\documentclass[reprint, showpacs]{revtex4-1}
\usepackage{graphicx}
\usepackage{amsfonts}
\usepackage{natbib}
\usepackage{float}

\newcommand{\be}{\begin{equation}}
\newcommand{\ee}{\end{equation}}
\newcommand{\bea}{\begin{eqnarray}}
\newcommand{\eea}{\end{eqnarray}}

\newcommand{\ep}{\varepsilon}
\newcommand{\al}{\alpha}
\newcommand{\bet}{\beta}
\newcommand{\mA}{\left< A \right>}

\begin{document}
\title{The distribution of localization measures of chaotic eigenstates in the stadium billiard}

\author{Benjamin Batisti\'c}
\author{\v Crt Lozej}
\author{Marko Robnik}

\affiliation{CAMTP - Center for Applied Mathematics and Theoretical
Physics, University of Maribor, Mladinska 3, SI-2000 Maribor, Slovenia, European Union}

\date{\today}

\begin{abstract}
  The localization measures $A$ (based on the information entropy) of localized
  chaotic eigenstates in the Poincar\'e-Husimi representation have a
  distribution on a compact interval $[0,A_0]$, which is well approximated
  by the {\em beta distribution}, based on our extensive numerical
  calculations. The system under study is the Bunimovich' stadium billiard,
  which is a classically ergodic system, also
  fully chaotic (positove Lyapunov exponent), but in the regime of a
  slightly distorted circle billiard (small shape parameter $\ep$)
  the diffusion in the momentum
  space is very slow. The parameter $\alpha=t_H/t_T$, where $t_H$ and
  $t_T$ are the Heisenberg time and the classical transport time (diffusion
  time),  respectively, is the important control parameter of the system,
  as in all quantum systems with the discrete energy spectrum. The
  measures $A$ and their distributions have been calculated for a
  large number of $\ep$ and eigenenergies.
  The dependence of the standard deviation $\sigma$ on $\al$
  is analyzed, as well as on the spectral parameter $\bet$ (level
  repulsion exponent of the relevant Brody level spacing distribution).
  The paper is a continuation of our recent paper (B. Batisti\'c, \v C. Lozej
  and M. Robnik, Nonlinear Phenomena in  Complex Systems {\bf 21}, 225 (2018)),
  where the spectral statistics and validity of the Brody level
  spacing distribution has been studied for the same system, namely
  the dependence of $\bet$ and of the mean value $<A>$ on $\al$. 
\end{abstract}

\pacs{01.55.+b, 02.50.Cw, 02.60.Cb, 05.45.Pq, 05.45.Mt}

\maketitle

\section{Introduction}
\label{sec1}

Quantum chaos (or more generally, wave chaos) is the study of phenomena in
the quantum domain, which correspond to the classical chaos. Thus, in the short
wavelength approximation we consider the dynamics of rays as the lowest order
approximation of the solution of the underlying wave equation, while in
the next order we have to consider the wave nature of the solutions, describing
the interference effects. The classical-quantum correspondence is thus, for example, 
entirely analogous to the correspondence between the Gaussian ray optics and
the solutions of the Maxwell equation as the  governing wave equation. The
major technique to bridge the classical and quantum phenomena is the
semiclassical mechanics. For an introduction to quantum chaos
see the books by St\"ockmann \cite{Stoe} and Haake \cite{Haake}, and a recent
review \cite{Rob2016}.

The quantum localization (or dynamical localization) of classical chaotic
diffusion in the time-dependent domain is one of the most important fundamental
phenomena in quantum chaos, discovered and studied first in the quantum 
kicked rotator \cite{Cas1979,Chi1981,Chi1988,Izr1990} by Chirikov, Casati,
Izrailev, Shepelyansky, Guarneri and many others, as an example of a
time-periodic Floquet system, whose  behaviour is quite typical.  See also
papers by Izrailev \cite{Izr1988,Izr1989} and his review \cite{Izr1990}.
Intuitively and qualitatively, the quantum diffusion does follow the
classical chaotic diffusion, but only up to the Heisenberg time (also called
break time), where it stops due to the (typically destructive) interfercence
effects. The Heisenberg time $t_H=2\pi\hbar/\Delta E$, where $\Delta E$ is
the mean energy level spacing (reciprocal energy level density), is an
important time scale in any quantum system with the discrete energy
spectrum. It is the time scale up to which the discreteness of the
evolution operator is not resolved. Note that $t_H$ and $\Delta E$ are
related through a Fourier transform of the evolving wave functions.

In the time-independent domain the quantum localization 
is manifested in the localized chaotic eigenstates. In the case of
the quantum kicked rotator, for example, one sees the exponentially
localized eigenstates in the dimensionless space of the angular momentum
quantum number. For an extensive review see \cite{Izr1990}.
This phenomenon is closely related to the Anderson
localization in one dimensional disordered lattices as shown
for the first time by Fishman, Grempel and Prange \cite{FGP1982},
and later discussed and studied by many others \cite{Stoe,Haake}.

Billiards are very convenient model systems, as they are simple
but nevertheless exhibit all generic properties of chaotic Hamiltonian
systems. The dynamical localization in billiards has been
reviewed by Prosen \cite{Pro2000}. We study the localization
properties of the chaotic eigenstates, which means studying the
structure of the Wigner functions (which are real but not positive
definite) or better the Husimi functions (which are real and positive
definite). The latter ones can be considered as a probability 
density. The separation of chaotic and regular eigenstates
is done by comparing the classical phase space with the structure
of their Wigner or Husimi functions.
The control parameter governing the degree of quantum localization is

\be  \label{defalpha}
\al = \frac{t_H}{t_T}
\ee
where $t_T$ is the dominating classical transport time (or diffusion time,
or ergodic time).
Batisti\'c and Robnik \cite{BatRob2013A} have recently studied the
localization of chaotic eigenstates in the mixed-type billiard
\cite{Rob1983,Rob1984}, after the separation of the
chaotic and regular eigenstates based on such quantum-classical
correspondence \cite{BatRob2013B}. Two localization
measures have been introduced, one based on the information entropy
denoted by $A$ and used in this paper, and the other one $C$ based
on the correlations. They have shown that $A$ and $C$ are linearly
related and thus equivalent, which confirms that the definitions are
physically sound and useful.

In a recent paper \cite{BLR2018} we have studied the localization
properties of chaotic eigenstates in the stadium billiard of
Bunimovich \cite{Bun1979}, which is ergodic and chaotic system
(positive Lyapunov exponents).
Studies of the slow diffusive regime in this system and the related
quantum localization were initiated in Ref. \cite{BCL1996},
while the detailed aspects of classical diffusion have been
investigated in our recent paper \cite{LozRob2018A}, where the
classical diffusion has been analyzed in detail, determining the important
classical transport time (diffusion time) $t_T$.

Another fundamental phenomenon in quantum chaos in the time-independent
domain is the statistics of the fluctuations in the energy spectra, which
are universal \cite{Stoe,Haake,Mehta,GMW,Rob1998} for classically
fully chaotic ergodic systems (described by the random matrix theories)
and for integrable systems (Poissonian statistics).
For this to apply one must be in
the sufficiently deep semiclassical limit (when $\al$ is large enough,
$\al \gg 1$, which can always be achieved by sufficiently small
effective $\hbar$). In the  general mixed type systems,
in the sufficiently deep semiclassical limit,
the spectral statistical properties
are determined solely by the type of classical motion,
which can be either regular or chaotic
\cite{Percival1973,BerRob1984,Rob1998,BatRob2010,BatRob2013A,BatRob2013B}.
The level statistics is Poissonian if the underlying classical invariant
component is regular. For chaotic extended states
the Random Matrix Theory (RMT) applies \cite{Mehta}, 
specifically the Gaussian Orthogonal Ensemble statistics
(GOE) in case of an antiunitary symmetry. This is the {\em Bohigas-Giannoni-Schmit
conjecture} \cite{Cas1980, BGS1984}, which has been proven only recently 
\cite{Sieber,Mueller1,Mueller2,Mueller3,Mueller4} using the semiclassical methods,
the periodic orbit theory developed around 1970 by Gutzwiller
(\cite{Gutzwiller1980} and the references therein), an approach initiated by 
Berry \cite{Berry1985}, well reviewed in \cite{Stoe,Haake}.

The classification regular-chaotic can be done by analyzing the
structure of eigenstates in the quantum phase space, based on the
Wigner functions, or Husimi functions \cite{BatRob2013B}. Of course, in the stadium
billiard all eigenstates are of the chaotic type, but can be strongly localized
if $\al$ is small enough, $\al\ll 1$.

The most important spectral statistical measure
is the level spacing distribution $P(S)$, assuming spectral
unfolding such that $\left<S\right>=1$. For integrable 
systems  and regular levels of mixed type systems
$P(S)=\exp\left(-S\right)$, whilst for extended chaotic systems it
is well approximated by the Wigner distribution 
$P(S)= \frac{\pi S}{2}\exp\left(-\frac{\pi}{4}\,S^2\right)$.
The distributions differ significantly in a small $S$ regime, where there
is no level repulsion in a regular system and a linear level repulsion,
$P(S)\propto S$, in a chaotic system. Localized chaotic states exhibit 
the fractional power-law level repulsion $P(S)\propto S^\beta$, as clearly
demonstrated recently by Batisti\'c and Robnik \cite{BatRob2010,BatRob2013A,BatRob2013B}.

The weak ($\beta<1$) level repulsion of localized chaotic states is empirically
observed, but the whole distribution $P(S)$ is globally theoretically not known.
Several different distributions which would extrapolate the small
$S$ behaviour were proposed. The most popular are the Izrailev
distribution \cite{Izr1988,Izr1989,Izr1990} and the Brody distribution
\cite{Bro1973,Bro1981}. 
The Brody distribution is a simple generalization of the Wigner distribution. 
Explicitly, the Brody distribution is

\be \label{BrodyP}
P_B(S) = c S^{\beta} \exp \left( - d S^{\beta +1} \right), \;\;\; 
\ee
where 

\be \label{Brodyab}
c = (\beta +1 ) d, \;\;\; d  = \left( \Gamma \left( \frac{\beta +2}{\beta +1}
 \right) \right)^{\beta +1}
\ee
with  $\Gamma (x)$ being the Gamma function. It interpolates the
exponential and Wigner distribution as $\beta$ goes from $0$ to $1$. 
 One  important theoretical plausibility argument by Izrailev in support of
such intermediate level spacing distributions is that
the joint level distribution of Dyson circular ensembles can be extended
to noninteger values of the exponent $\beta$ \cite{Izr1990}.
The Izrailev distribution is a bit more complicated but has the feature of being
a better approximation for the GOE distribution at $\beta=1$.
However, recent numerical results show that Brody distribution 
is slightly better in describing real data
\cite{BatRob2010,BatRob2013A,ManRob2013,BatManRob2013}, 
and is simpler, which is the reason why we prefer and use it.

In the previous paper \cite{BLR2018} it has been shown
that there is a {\em linear} functional
relation between the level repulsion parameter $\beta$ and the mean
localization  measure $<A>$ in the stadium billiard, in analogy with the
quantum kicked rotator, but different from  the above mentioned
mixed-type billiard. Also, $\beta$ is a unique function
of $\al$, which has been discussed for the first time by Izrailev
\cite{Izr1988,Izr1989,Izr1990},
where he numerically studied the quantum kicked rotator.
His result showed that the parameter $\beta$, which
was obtained using the Izrailev distribution, is functionally related to the
localization measure defined by the information entropy of the
eigenstates in the angular momentum representation. His results were
recently confirmed and extended, with the much greater numerical accuracy
and statistical significance \cite{ManRob2013,BatManRob2013}.
Moreover, in Ref. \cite{BatRob2013A} it has been demonstrated that
$\beta$ is a unique function of $\mA$ in the billiard with the mixed phase space
\cite{Rob1983,Rob1984}, but is not linear. Finally, Manos and Robnik
\cite{ManRob2015}, have observed that the localization measure
in the quantum kicked rotator has a nearly Gaussian distribution,
and this was a
motivation for the work in the present paper, where we study
the distributions of $A$ in the stadium billiard, systematically
in almost all regions of interest
(determined by the shape parameter $\ep$ and the energy), and also
the dependence of its standard deviation on the control parameter
$\al$, while the dependence of $<A>$ on $\al$ as an empirical
rational function is already known from our previous paper \cite{BLR2018}.

The paper is organized as follows. In section \ref{sec2} we define the system
and the Poincar\'e - Husimi functions. In section \ref{sec3} we show
examples of Poincar\'e - Husimi functions and calculate
the moments of the distributions of the localization measures $P(A)$.
In section \ref{sec4} we analyze the distributions
$P(A)$ extensively and in detail, and demonstrate that they are
very well decsribed by the beta distribution. In section \ref{sec5}
we consider the implications of localization for the statistical
properties of the energy spectra, in particular for the level
spacing distribution. In section \ref{sec6} we conclude and discuss
the main results.

\section{The billiard systems and definition of the Poincar\'e-Husimi functions}
\label{sec2}

\begin{figure}[H]
  \centering
  \includegraphics{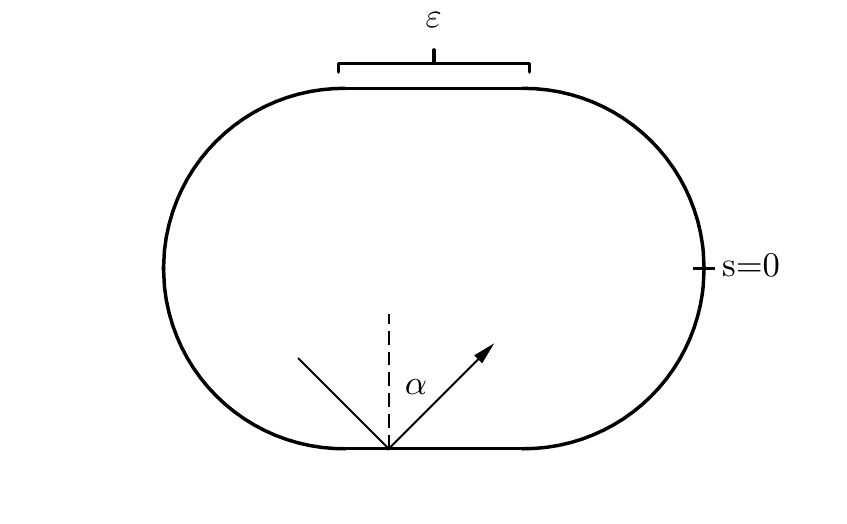}
  \caption{The geometry and notation of the stadium billiard of Bunimovich.}
  \label{figlr1}
\end{figure}

The (shape of the) stadium billiard ${\cal B}$ of Bunimovich
\cite{Bun1979} is defined as two semicircles of
radius 1 connected by two parallel straight lines of length $\ep$, as shown
in Fig. \ref{figlr1}. We study the dynamics of a point particle
moving freely inside the billiard, and experiencing specular
reflection when hitting the boundary. In this section we follow our previous
paper \cite{BLR2018} and go further.

For a 2D billiard the most natural coordinates in the phase space
$(s,p)$ are the arclength $s$ round the billiard boundary, $s\in [0,{\cal L}]$,
where ${\cal L}$ is the circumference, and the sine of
the reflection angle, which is the component of the unit velocity
vector tangent to the boundary at the collision point, equal to $p=\sin \al$,
which is the canonically conjugate momentum to $s$. These are the
Poincar\'e-Birkhoff coordinates. The bounce map $(s_1,p_1)
\rightarrow (s_2,p_2)$ is area preserving \cite{Berry1981}, and the phase portrait
does not depend on the speed (or energy) of the particle. Quantum mechanically
we have to solve the stationary Schr\"odinger equation, which in a
billiard is just the Helmholtz equation 

\be \label{Helmholtz}
\Delta \psi + k^2 \psi =0
\ee
with the Dirichlet boundary conditions  $\psi|_{\partial {\cal B}}=0$.
The energy is $E=k^2$. The important quantity is
the boundary function 

\be  \label{BF}
u(s) = {\bf n}\cdot \nabla_{{\bf r}} \psi \left({\bf r}(s)\right),
\ee
which is the normal derivative of the wavefunction $\psi$ at the 
point $s$ (${\bf n}$ is the unit outward normal vector). 
It satisfies the integral equation

\be \label{IEBF}
u(s) = -2 \oint dt\; u(t)\; {\bf n}\cdot\nabla_{{\bf r}} G({\bf r},{\bf r}(t)),
\ee
where $G({\bf r},{\bf r'}) = -\frac{i}{4} H_0^{(1)}(k|{\bf r}-{\bf r'}|)$ is
the Green function in terms of the Hankel function $H_0(x)$. It is important
to realize that the boundary function $u(s)$ contains complete information
about the wavefunction at any point ${\bf r}$ inside the billiard by the equation

\be \label{utopsi}
\psi_m({\bf r})  = - \oint dt\; u_m(t)\; G\left({\bf r},{\bf r}(t)\right).
\ee
Here $m$ is just the index (sequential quantum number) of the $m$-th eigenstate.
Now we go over to the quantum phase space. We can calculate the Wigner
functions \cite{Wig1932} based on $\psi_m({\bf r})$. However, in billiards it is advantageous to
calculate the Poincar\'e - Husimi functions. The Husimi functions \cite{Hus1940} are
generally just Gaussian smoothed Wigner functions. Such smoothing makes
them positive definite, so that we can treat them somehow as quasi-probability 
densities in the quantum phase space, and at the same time we eliminate the
small oscillations of the Wigner functions around the zero level, which do
not carry any significant physical contents, but just obscure the picture.
Thus, following  Tualle and Voros \cite{TV1995} and B\"acker et al
\cite{Baecker2004}, we introduce \cite{BatRob2013A,BatRob2013B}. 
the properly ${\cal L}$-periodized coherent states
centered at $(q,p)$, as follows

\bea \label{coherent}
c_{(q,p),k} (s) & =  & \sum_{m\in {\bf Z}} 
\exp \{ i\,k\,p\,(s-q+m{\cal L})\}  \times \\ \nonumber
 & \exp & \left(-\frac{k}{2}(s-q+m{\cal L})^2\right). 
\eea
The Poincar\'e - Husimi function is then defined as the absolute square
of the projection of the boundary function $u(s)$ onto the coherent
state, namely

\be \label{Husfun}
H_m(q,p) = \left| \int_{\partial {\cal B}} c_{(q,p),k_m} (s)\;
u_m(s)\; ds \right|^2.
\ee
The {\em entropy localization measure} of a {\em single
eigenstate}  $H_m(q,p)$, denoted by $A_m$ is defined as

\be \label{locA}
A_m = \frac{\exp I_m}{N_c},
\ee
where

\be  \label{entropy}
I_m = - \int dq\, dp \,H_m(q,p) \ln \left((2\pi\hbar)^f H_m(q,p)\right)
\ee
is the information entropy.  Here $f$ is the number of degrees
of freedom (for 2D billiards $f=2$, and for surface of section it is
$f=1$) and $N_c$ is a number of cells on the 
classical chaotic domain, $N_c=\Omega_c/(2\pi\hbar)^f$, where
$\Omega_c$ is the classical phase space volume of the classical chaotic component.
In the case of the
uniform distribution (extended eigenstates) $H=1/\Omega_C={\rm const.}$
the localization measure is $A=1$, while in the case of the strongest localization
$I=0$, and $A=1/N_C \approx 0$.
The Poincar\'e - Husimi function $H(q,p)$
(\ref{Husfun}) (normalized) was calculated on the grid points $(i,j)$
in the phase space $(s,p)$,  and
we express the localization measure in terms of the discretized function.
In our numerical calculations we have put $2\pi\hbar=1$, and
thus we have $H_{ij}=1/N$, where $N$ is the number of grid points,
in case of complete extendedness, while for maximal localization
we have $H_{ij}=1$ at just one point, and zero elsewhere.
In all calculations
have used the grid of $400\times 400$ points, thus $N = 160000$.

\begin{figure}[H]
  \centering
  \includegraphics[width=9cm]{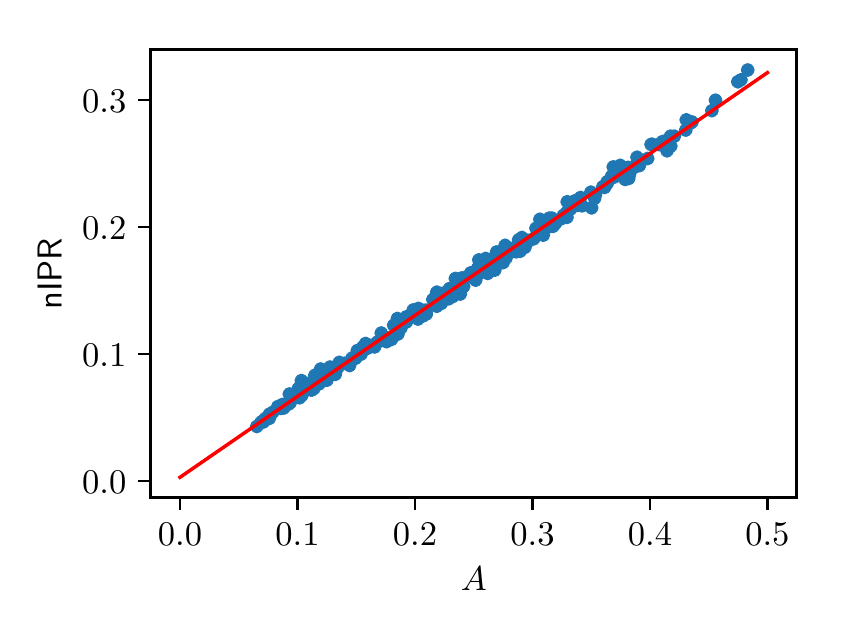}
  \caption{The normalized inverse participation ratio as a localization measure,
    as a function of $A$. They are linearly related and thus equivalent.
    The slope is 0.64.} 
  \label{nIPRofA}
\end{figure}

As mentioned in the introduction, the definition of localization measures
can be diverse, and the question arises to what extent are the results
objective and possibly independent of the definition. Indeed, in reference
\cite{BatRob2013A}, it has been shown that $A$ and $C$ (based on the corelations)
are linearly related and thus equivalent. Moreover, we have introduced also
the normalized inverse participation ratio $nIPR$, defined as follows

\be \label{nIPR}
nIPR = \frac{1}{N} \frac{1}{\sum_{i,j} H_{ij}^2},
\ee
for each individual eigenstate $m$. However, because we expect fluctutaions
of the localization measures even in the quantum ergodic regime (due to the scars etc),
we must perform some averaging over an ensemble of eigenstates, and for this we have
chosen $100$ consecutive eigenstates. Then, by doing this for all possible data for the
stadium at various $\ep$ and $k$, we ended up with the perfect result that
the $nIPR$ and $A$ are linearly related and thus also equivalent, as shown in
Fig. \ref{nIPRofA}. In the following we shall use exclusively $A$ as the measure
of localization.

The central object of interest in this paper is the distribution $P(A)$
of the localization measures $A_m$ within a certain interval of 1000 consecutive
eigenstates indexed by $m$, around some central value $\bar{k}$.
We have done this for 17 different values of $\ep$ and for each $\ep$
for 12 different values of $\bar{k}$.
Each distribution function $P(A)$, generated by the segment of
1000 consecutive values $A_m$, is defined on a compact interval $[0,A_0]$.
Ideally, according to the above Eqs. (\ref{locA},\ref{entropy}), the maximum
value of $A$ should be $1$, if the Husimi function were entirely and uniformly
extended. However, this is never the case, as the Husimi functions have zeros
and oscillations, and thus we must expect a smaller maximal value, smaller  than $1$,
which in addition might vary from case to case, depending on $k$ and the grid size.
So long as we do not have a theoretical prediction for $A_0$, we must proceed
empirically. Therefore we have checked several values of $A_0$ around $A_0=0.7$,
and found that the latter value is the best according to several criteria.
See also the discussion at the end of section \ref{sec4}.

We shall look at the moments of $P(A)$, namely

\be \label{locA2}
\left< A \right>= \int_0^{A_0}A\, P(A) \,dA, \,\,\,
\left< A^2 \right>=  \int_0^{A_0}A^2\, P(A) \,dA,
\ee
and the standard deviation 

\be \label{locA3}
\sigma =\sqrt{\left< A^2 \right> - \left< A \right>^2}.
\ee
For the numerical calculations of the eigenfunctions $\psi_m({\bf r})$
and the corresponding energy levels $E_m=k_m^2$  we have used the
Vergini-Saraceno method  \cite{VerSar1995}. Also, we have calculated only the
odd-odd symmetry class of solutions.

\section{ Moments of A  and examples of Poincar\'e - Husimi functions}
\label{sec3}

The system parameter governing the localization phenomenon
$\al = t_H/t_T$, as introduced in Eq. (\ref{defalpha}),
in a quantum billiard described by the Schr\"odinger equation
(Helmholtz equation) Eq. (\ref{Helmholtz}), becomes

\begin{equation} \label{A7}
\alpha = \frac{2k}{N_T}.
\end{equation}
where $N_T$ is the discrete classical transport time, that is the
characteristic number of collisions of the billiard particle
necessary  for the global spreading of the ensemble of uniform in $s$
initial points (excluding the bouncing ball intervals)
at zero momentum in the momentum space.
This quantity $N_T$ can be defined in various ways as discussed
in references \cite{BLR2018,BatRob2013A,BatRob2013B}, where
the derivation of $t_T$, $N_T$ and $\al$ is given. It is shown
there that $N_T\propto \ep^{-2.3}$ for small $\ep\le 0.1$.

The condition for the occurrence of dynamical localization
$\alpha \le 1$ is now expressed in the inequality

\begin{equation} \label{A8}
k \le \frac{N_T}{2},
\end{equation}
although the empirically observed transitions are not at all sharp with $\al$.
More precisely,  as in Ref. \cite{BLR2018},
$t_T$ is defined as the time at which an ensemble of  initial conditions in
the momentum space with initial Dirac delta distribution with
zero variance reaches a certain fraction of the asymptotic value.
In the stadium billiard for small $\ep$ we have a
diffusive regime and thus $t_T$ can be defined
as the diffusion time extracted from the exponential approach of
the momentum variance to its asymptotic value $1/3$,
as has been recently carefully studied in Ref. \cite{LozRob2018A}.
In ref. \cite{BLR2018} (see Table I) we have published the values
of $N_T$ for the stadium billiard, for 40 different values of
the shape parameter $\ep$, for the criteria $50\%, 70\%,80\%,90\%$ of the asymptotic
value of the momentum variance $\left< p^2 \right>$ and for the
exponential model.

\begin{figure}[H]
  \centering
  \includegraphics[width=9cm]{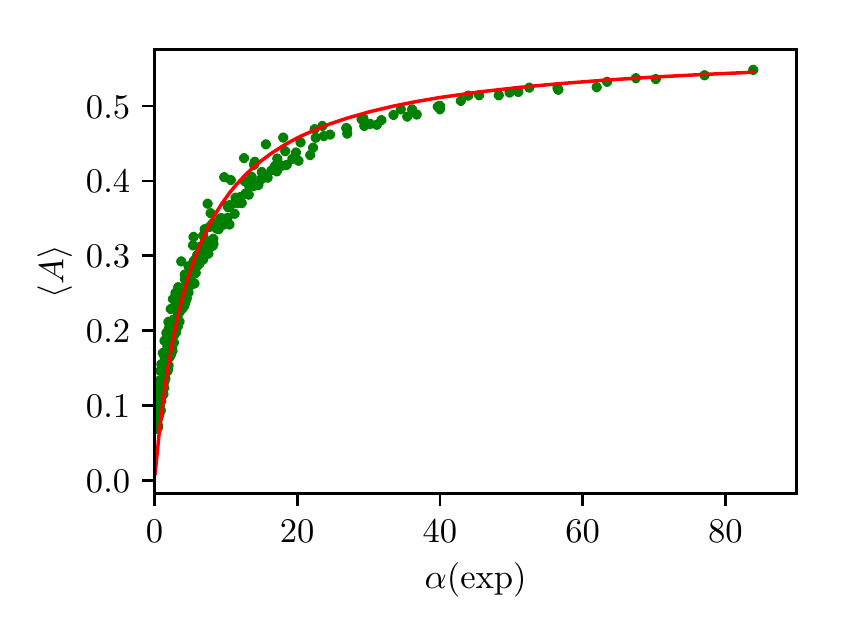}
  \caption{The entropy localization measure $\mA$ as a function of $\al$ fitted
  by the function (\ref{Avsalphaeq}), based on $N_T$ from the exponential
  diffusion law (see Table I in ref. \cite{BLR2018}), with
  $A_{\infty}=0.58$ and $s=0.19$.} 
  \label{Avsalphaexp}
\end{figure}

In Fig. \ref{Avsalphaexp}  we show the dependence of $\mA$ on $\al$,
where $\al$ is calculated using $N_T$ from the expnential law. The transition
from strong localization of small $\mA$ and $\al$ to complete delocalization
$\al \gg 1$ is quite smooth, over almost two decadic orders of magnitude.
As we see, $\mA$ is well fitted by a rational function of $\al$,  namely

\be  \label{Avsalphaeq}
\mA = A_{\infty} \frac{s\al}{1 +s \al},
\ee
where the values of the two parameters are $A_{\infty}=0.58$ and $s=0.19$.

\begin{figure}[H]
  \centering
  \includegraphics[width=9cm]{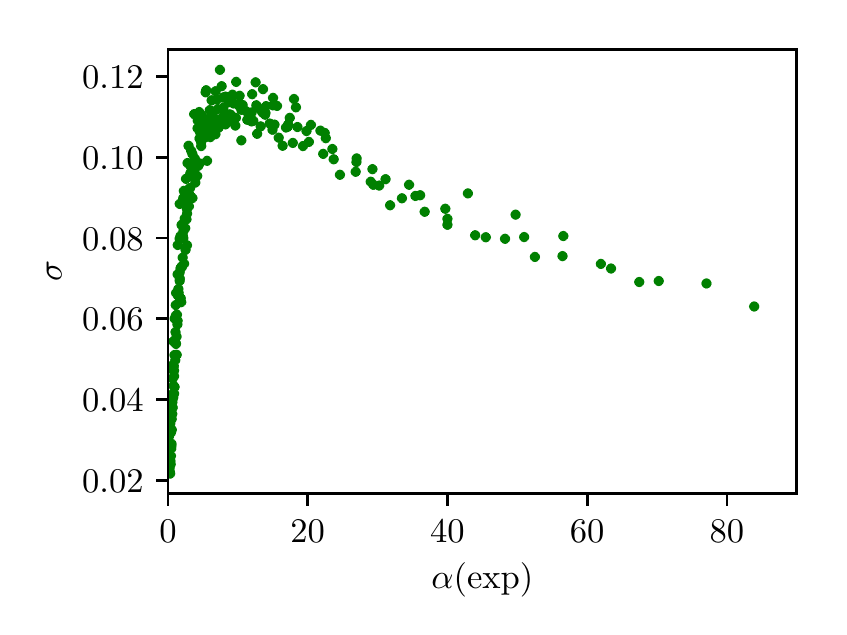}
  \caption{The standard deviation $\sigma$ as a function of the
    $\al$ for variety of stadia of
    different shapes $\ep$ and energies $E=k^2$, as defined in the text.}
  \label{sigmaofalpha}
\end{figure}

In Fig. \ref{sigmaofalpha} we show the dependence of $\sigma$ defined in
Eq. (\ref{locA3}) upon $\al$ also using the exponential model for $N_T$.
The results are functionaly the same when using the other definitions
of $N_T$, so that we do not show them here.  

We see that while $\mA$ is
a monotonically increasing function of $\al$ (Fig. \ref{Avsalphaexp}),
the standard deviation $\sigma$ starts at zero,
is small for small $\al$, but rises sharply,
and reaches some maximum at about $\al\approx 10$, and then decreases
very slowly at large values of $\al$. Thus both, the very strongly
localized eigenstates, mimicking invariant tori, and the
entirely delocalized (ergodic) eigenstates have 
small spreading $\sigma$ around the mean value $\mA$. According to the
quantum ergodic theorem of Shnirelman \cite{Shnirelman1974}
$\sigma$ should tend to zero when $\al\rightarrow \infty$,
and rescaled $\mA\rightarrow 1$, but the transition
to that regime might be very slow as suggested by Fig. \ref{sigmaofalpha}.
However, it is very difficult to judge this quantitatively,
as at large $\al$ we have very few physically reliable data
points, so it is too early to draw any definite conclusion
about the asymptotic behavior at $\al \rightarrow \infty$.
More numerical efforts are needed, currently not feasible.

\begin{figure*}
  \begin{centering}
    \includegraphics[width=1\textwidth]{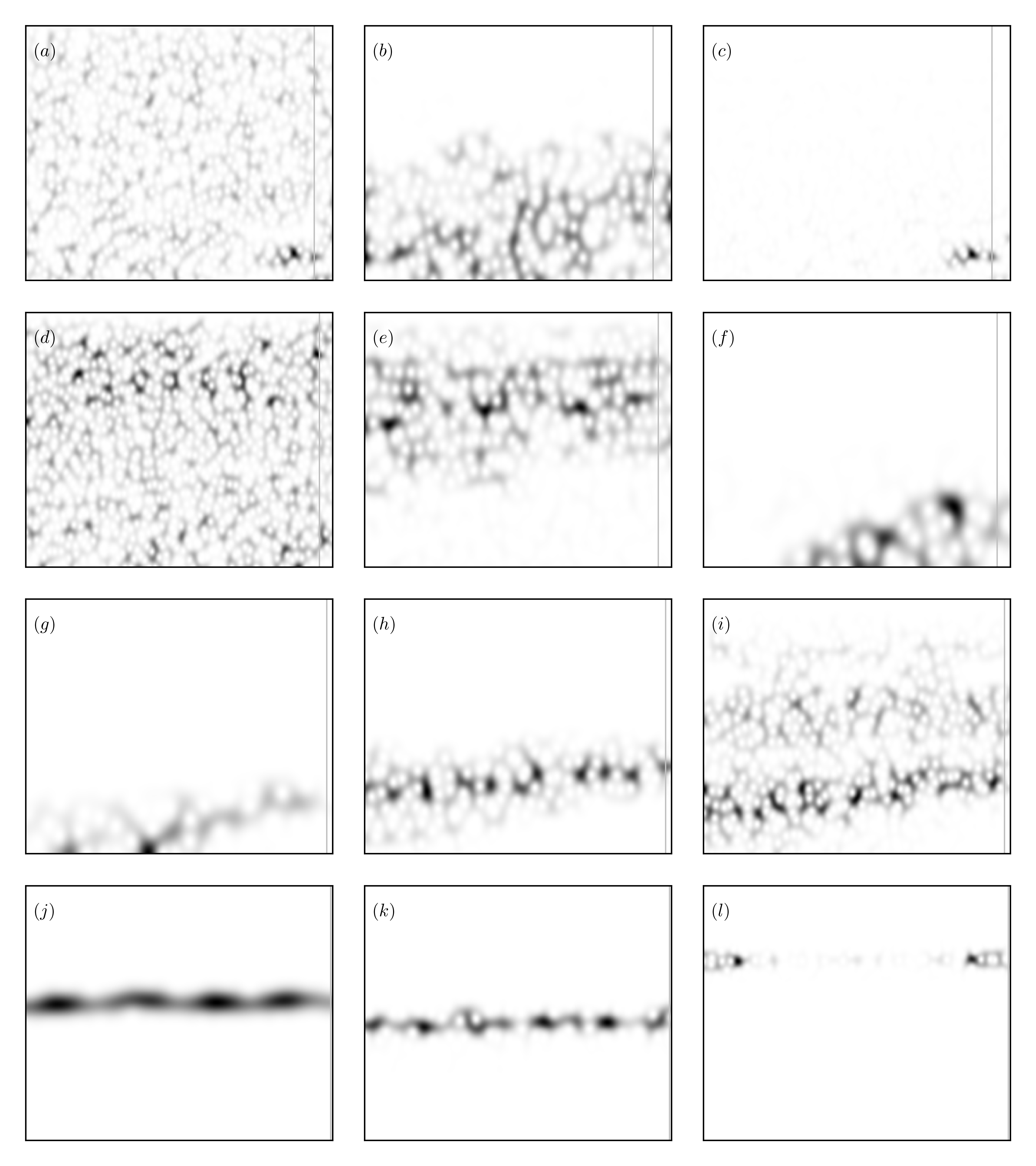}
    \par\end{centering}
  \caption{We show plots of Poincar\'e - Husimi functions for a
    representative selection of eigenstates:
    Row (a-c): $\ep=0.2$, $k$: $3720.01170174$, $1480.01417766$ , $3720.01594071$;
    Row (d-f): $\ep=0.14$, $k$: $3720.01379303$, $1480.03480316$, $640.05384413$;
    Row (g-i): $\ep=0.06$, $k$: $640.05384413$, $1480.05082005$, $2600.01768874$;
    Row (j-l): $\ep=0.02$, $k$: $640.00758741$, $1480.03406374$,  $3720.00973621$.
    The thin vertical line denotes the size of the interval $\ep/2$.
  }
  \label{Husimipic}
\end{figure*}

The Poincar\'e - Husimi functions describe the structure of the localized
chaotic eigenstates. In Fig. \ref{Husimipic} we show some selection of
typical Poincar\'e - Husimi functions for various values of $\ep$ and $k$,
and the corresponding $\al$. We show only the upper right quadrant 
$s \in [0,{\cal L}/4]$, $p\in [0,1]$  of the
classical phase space, as due to the symmetries (two reflection symmetries
and the time reversal symmetry) all four quadrants are equivalent.

At large $\ep$ and fixed $k$, we have small small $N_T$ and according
to Eq.(\ref{A7}) $\al \gg 1$, we observe mainly ergodic eigenstates,
in agreement with the quantum ergodic theorem \cite{Shnirelman1974}, that is
fully extended states, exemplified in (a). Nevertheless, there are some exceptions,
asymptotically of measure zero, where we observe partial localization,
as shown in (b). Moreover, there can be strongly localized states corresponding
to the scaring around and along an unstable periodic orbit as exemplified
in (c) and (l). More precisely, the area of scars of eigenfunctions
$\psi_m({\bf r})$ goes to zero, and the relative number of scarred states
goes to zero as $\hbar\rightarrow 0$ or $m\rightarrow \infty$ \cite{Heller1984}.

As we decrease $\ep=0.14$ and $\al$, thereby increasing $N_T$, the degree of localization
increases, thus $\mA$ is decreasing as shown in (d-f). At still lower value of $\ep=0.06$
we see even more strongly localized states exemplified in (g-i). Finally, at the smallest
value of $\ep=0.02$ that we considered in our numerical calculations, we see only strongly
localized eigenstates mimicking invariant tori, in (j-l), although the system is
classically ergodic, but obviously is full of cantori with very low transport
permeability.

\section{The distributions of the localization measures $A$}
\label{sec4}

\begin{figure*}
  \begin{centering}
    \includegraphics[width=1\textwidth]{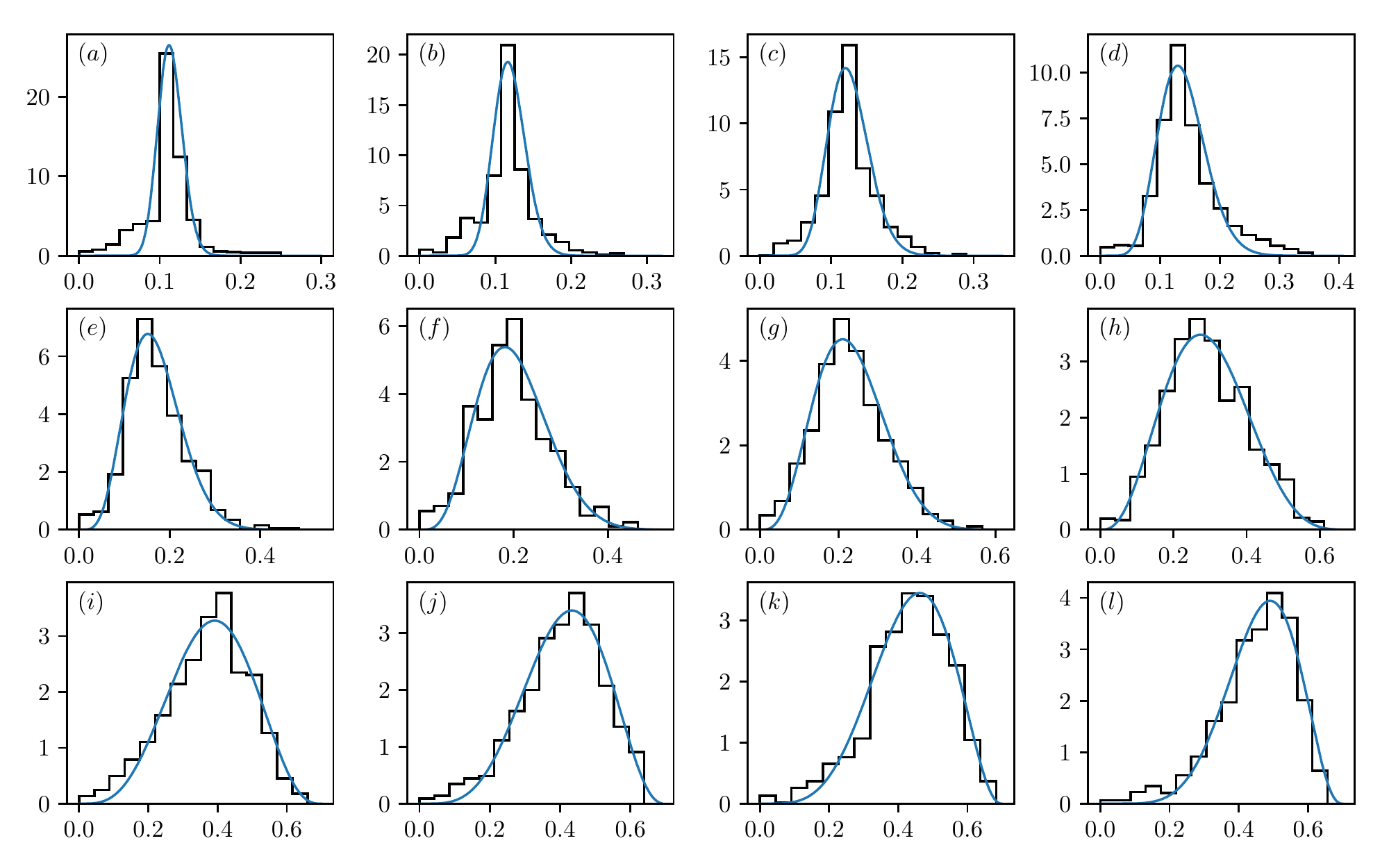}
    \par\end{centering}
  \caption{The distributions $P(A)$ of the entropy localization measure $A$
    for $k_0=640$ and various $\ep$ (from (a) to (l)):
    0.02,  0.03,  0.04,  0.05, 0.06,  0.07,  0.08,  0.1,
    0.14,  0.16, 0.18, 0.2.
    }
  \label{PA640}
\end{figure*}

\begin{figure*}
  \begin{centering}
    \includegraphics[width=1\textwidth]{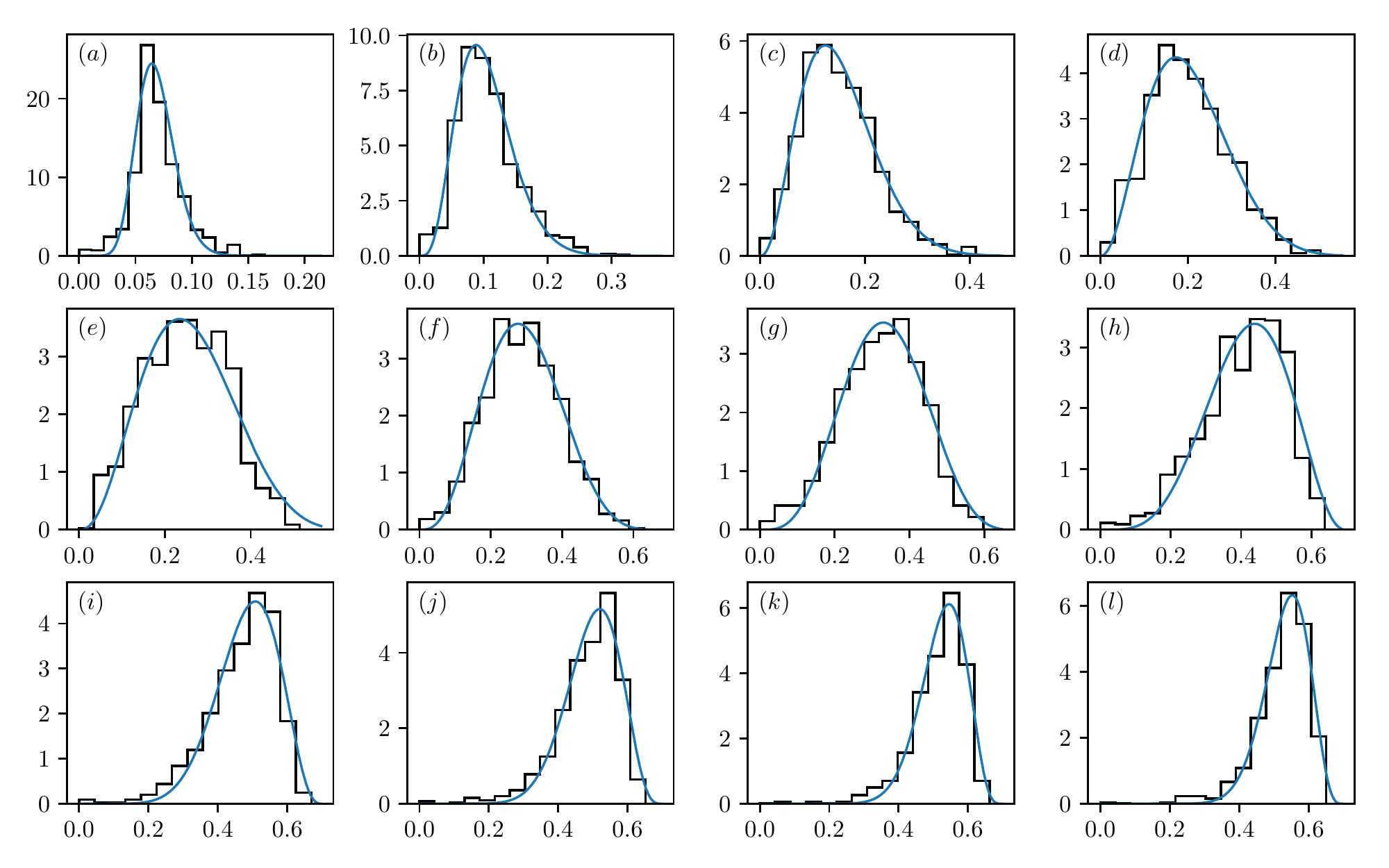}
    \par\end{centering}
  \caption{The distributions $P(A)$ of the entropy localization measure $A$
    for $k_0=2320$ and various $\ep$ (from (a) to (l)):
    0.02,  0.03,  0.04,  0.05, 0.06,  0.07,  0.08,  0.1,
    0.14,  0.16, 0.18, 0.2.
    }
  \label{PA2320}
\end{figure*}

\begin{figure*}
  \begin{centering}
    \includegraphics[width=1\textwidth]{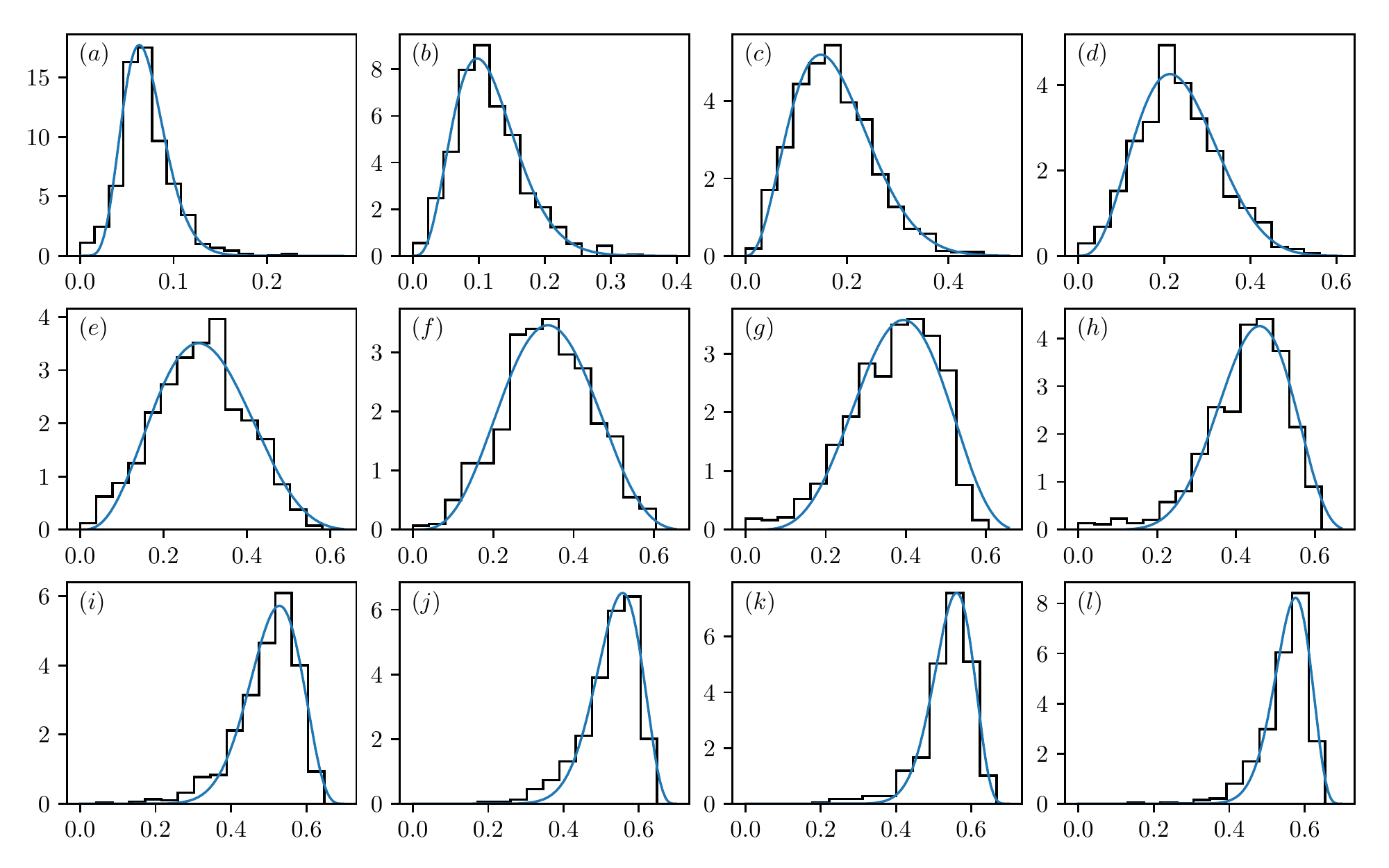}
    \par\end{centering}
  \caption{The distributions $P(A)$ of the entropy localization measure $A$
    for $k_0=3440$ and various $\ep$ (from (a) to (l)):
    0.02,  0.03,  0.04,  0.05, 0.06,  0.07,  0.08,  0.1,
    0.14,  0.16, 0.18, 0.2.
    }
  \label{PA3440}
\end{figure*}

In this section we present the central results of this paper, namely the
distribution functions of the localization measures $A$.  It turns
out that each distribution can be  very well characterized and described
by the so-called {\em beta distribution}

\be  \label{betadistr}
P(A) = C A^a (A_0-A)^b,
\ee
where $A_0$ is the upper limit of the interval $[0,A_0]$ on which $P(A)$ is defined,
in our case, as explained in Sec. \ref{sec2}, we have chosen $A_0=0.7$.
The two exponents $a$ and $b$ are positive  real numbers, while $C$ is the
normalization constant such that $\int_0^{A_0} P(A)\,dA = 1$, i.e.

\be \label{C}
C^{-1} = A_0^{a+b+1} B(a+1,b+1),
\ee
where $B(x,y) = \int_0^1 t^{x-1} (1-t)^{y-1} dt$ is the beta function.
Thus we have

\be \label{mA}
\mA = A_0 \frac{a+1}{a+b+3},
\ee
and for the second moment

\be \label{2mA}
\left< A^2 \right>  = A_0^2 \frac{(a+2)(a+1)}{(a+b+4)(a+b+3)}
\ee
and therefore for the standard deviation  $\sigma$  (\ref{locA3}),

\be \label{sigmaA}
\sigma^2  = 
A_0^2 \frac{(a+2)(b+2)}{(a+b+4)(a+b+3)^2}.
\ee
such that asymptotically  $\sigma \approx A_0 \frac{\sqrt{b+2}}{a}$ when
$a\rightarrow \infty$.
In the figures \ref{PA640}, \ref{PA2320}, \ref{PA3440}
we show a selection of typical distributions $P(A)$.
In all cases for $A_0$ we have chosen $A_0=0.7$. By $k_0$ we denote the starting
value of $k$ intervals on which we calculate the 1000 successive eigenstates.
It should be noted that the statistical significance is very high,
which has been carefully checked by using a (factor 2) smaller number of objects
in almost all histograms, as well as by changing the size of the boxes.

The limiting case $a \rightarrow \infty$ in Eqs.(\ref{mA},\ref{sigmaA}) comprising the
fully extended states in the limit $\al \rightarrow\infty$ shows that the
distribution tends to the Dirac delta function peaked at $A_0$,
thus $\sigma=0$ and $P(A)=\delta(A-A_0)$, in agreement with Shnirelman's theorem
\cite{Shnirelman1974}.

The figures clearly show that the fit by the beta distribution (\ref{betadistr})
is excellent, except for few cases at small $\ep$.
The corresponding values of $a$, $b$, and $\al$ are shown in
Table I. The qualitative trend from strong localization to weaker
localization or even complete extendedness (ergodicity) with increasing $k_0$
and $\ep$ is clearly visible.

\begin{figure*}
  \begin{centering}
    \includegraphics[width=1\textwidth]{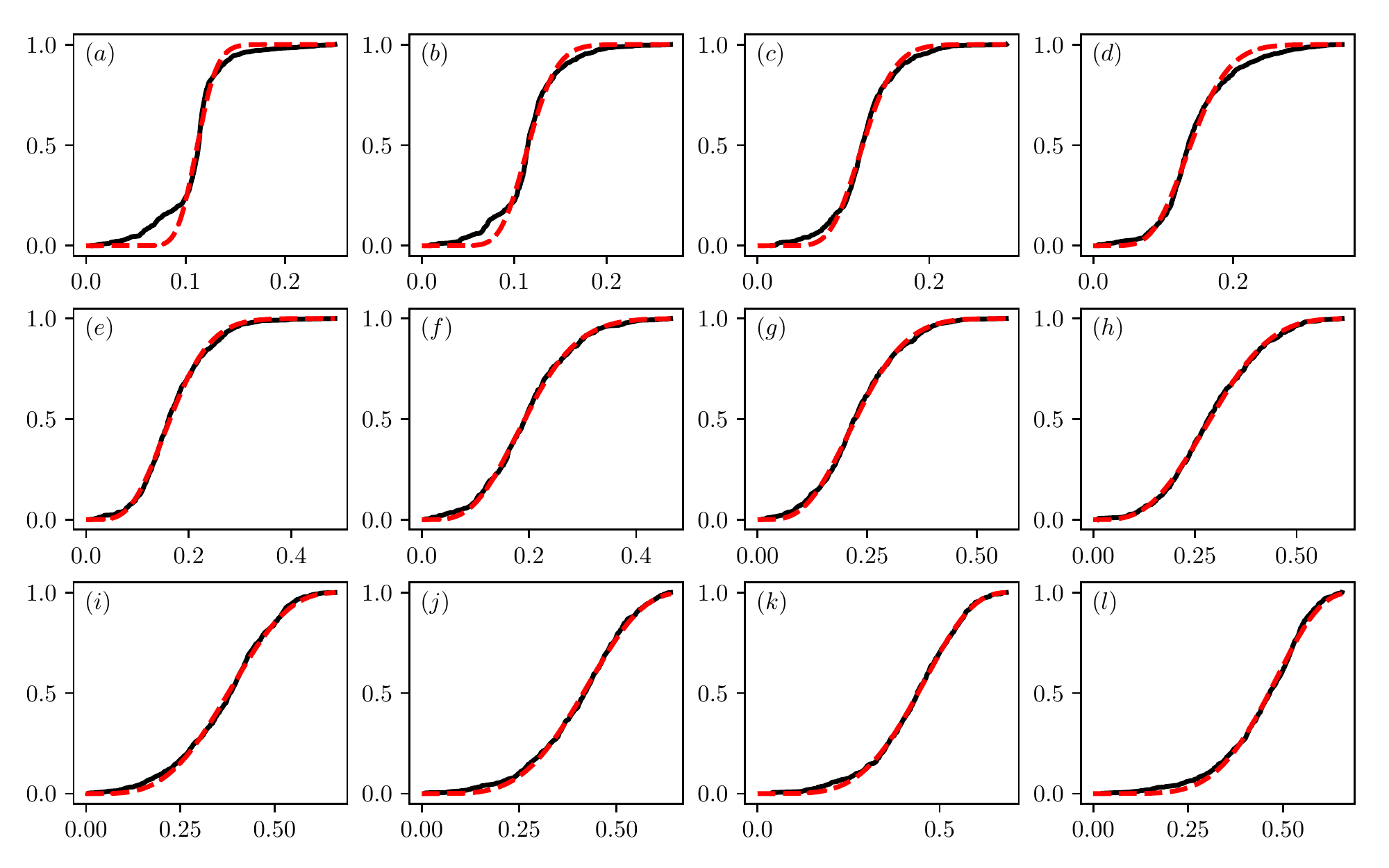}
    \par\end{centering}
  \caption{The cumulative distributions $W(A)$ of the entropy localization measure $A$
    for $k_0=640$ and various $\ep$ (from (a) to (l)):
    0.02,  0.03,  0.04,  0.05, 0.06,  0.07,  0.08,  0.1,
    0.14,  0.16, 0.18, 0.2. They correspond to Fig. \ref{PA640}.
    The numerical data are represented by the full line, the dashed one is
    the best fitting cumulative beta distribution.
    }
  \label{cumPA640}
\end{figure*}  
The agreement of the distributions $P(A)$, based on 1000 numerical values
per histogram, with the empirically found beta distribution, is demonstrated
for the cumulative distribution $W(A)$,

\be \label{cumPA}
W(A) = \int_0^A P(x)\, dx.
\ee
in Fig. \ref{cumPA640}, for $k_0=640$ and various $\ep$. The worst case is
at small $\ep=0.02$  (a), while the best one is for $\ep=0.2$ (l). The quality of the
fit for other values of $\ep$ and $k_0$ is much better, comparable to (l). Therefore we
may conclude that apart from the smallest $\ep=0.02$, the beta distribution is
the semiempirically right description of the distributions $P(A)$.

\begin{table}
  \center
\begin{tabular}{ | p{1.7cm} | p{1.7cm}  | p{1.7cm} | }
  \hline
  \multicolumn{3}{|c|}{Parameters of the beta distributions}\\
  \hline
  $a$  &           $b$  &   $\alpha$ \\
  \hline
 41.694174 &  222.486122 & 0.087593 \\\hline
 18.693580 &   97.355740 & 0.233279 \\\hline
 13.554691 &   66.554374 & 0.468864 \\\hline
  7.789110 &   33.474273 & 0.779063 \\\hline
  4.745791 &   17.092065 & 1.170018 \\\hline
  3.784283 &   11.147294 & 1.673203 \\\hline
  3.445179 &    8.128386 & 2.273535 \\\hline
  2.869704 &    4.433526 & 3.753666 \\\hline
  3.390150 &    2.806308 & 7.441860 \\\hline
  4.052971 &    2.581244 & 9.770992 \\\hline
  4.288946 &    2.245432 &12.549020 \\\hline
  5.352111 &    2.401925 &15.609756 \\\hline\hline
 12.767080 &  125.817047 & 0.317525 \\\hline
  3.519266 &   24.397953 & 0.845635 \\\hline
  2.567093 &   12.051933 & 1.699634 \\\hline
  2.397450 &    7.362461 & 2.824102 \\\hline
  2.570951 &    5.339960 & 4.241316 \\\hline
  3.184420 &    4.989456 & 6.065359 \\\hline
  3.553229 &    4.183009 & 8.241563 \\\hline
  3.894103 &    2.440766 &13.607038 \\\hline
  7.260777 &    2.884395 &26.976744 \\\hline
  9.534313 &    3.451923 &35.419847 \\\hline
 13.582757 &    4.012060 &45.490196 \\\hline
 14.385305 &    3.975161 &56.585366 \\\hline\hline
  6.324855 &   64.722649 & 0.470814 \\\hline
  3.518886 &   22.076947 & 1.253873 \\\hline
  2.668046 &   10.146521 & 2.520147 \\\hline
  3.149723 &    7.293627 & 4.187462 \\\hline
  3.000757 &    4.566317 & 6.288848 \\\hline
  3.604156 &    3.959419 & 8.993464 \\\hline
  4.354036 &    3.582313 &12.220249 \\\hline
  6.856759 &    3.788696 &20.175953 \\\hline
 11.037403 &    3.748447 &40.000000 \\\hline
 13.288212 &    3.566739 &52.519084 \\\hline
 19.143654 &    4.832500 &67.450980 \\\hline
 21.812546 &    4.992295 &83.902439 \\\hline
\end{tabular}
\caption{Parameters of the best fitting beta distributions
  of Figs. \ref{PA640}, \ref{PA2320}, \ref{PA3440}, successively.
$A_0$ is fixed, $A_0=0.7$.}
\end{table}

\begin{figure}[H]
  \centering
  \includegraphics[scale=0.75]{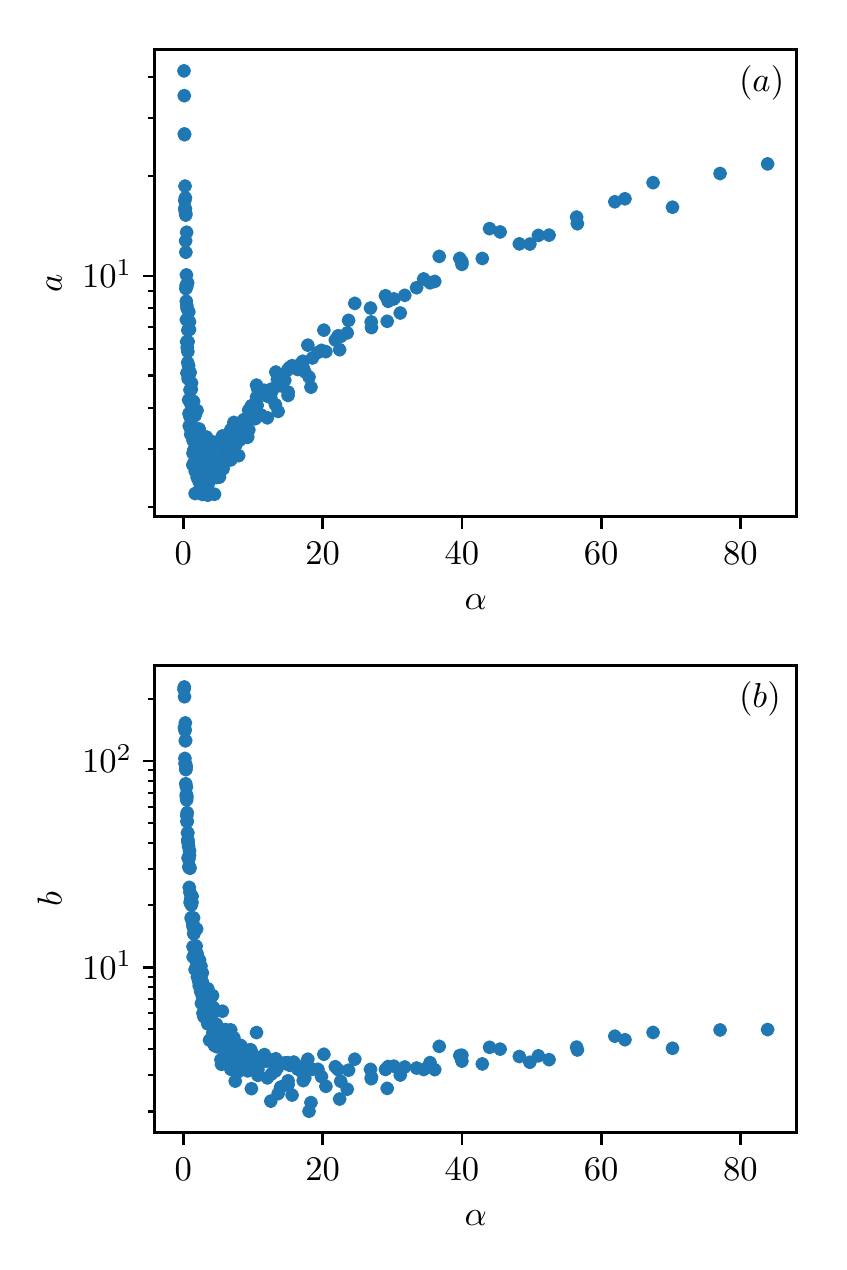}
  \caption{The plot of $a$ and $b$ versus $\al$.}
  \label{abofalpha}
\end{figure}

We should stress that there is of course some arbitrarines in defining $A_0$,
so long as we do not have a theoretical prediction for its value. 
So far we have taken $A_0=A_{{\rm max}}=0.7$, but nevertheless tried also the choice
of $A_0$ being the largest member $A$ in each histogram, and found no
significant qualitative changes, but only quite minor quantitative differences in
the fitting curves of the histograms. In both cases $a$ and $b$ are {\em not}
unique functions of $\al$, while $<A>$ and $\sigma$ are close to being
unique functions of $\al$ as demonstrated in Figs. \ref{Avsalphaexp} and
\ref{sigmaofalpha}, and approximated by a fit in Eq. (\ref{Avsalphaeq}).
In Fig. \ref{abofalpha} we show in log-lin plot of
$a$ and $b$ versus $\alpha$.

\section{Implications of localization for the spectral statistics}
\label{sec5}

To get a good estimate of $\beta$ we need many more levels (eigenstates)
than in calculating $\mA$. 
The parameter $\beta$ was computed for 40 different values of the parameter
$\epsilon$: $\epsilon_j = 0.02 + 0.005\,j$ where $j \in [0,1,\dots, 39]$ and on
12 intervals in $k$ space: $(k_i,k_{i+1})$ where $k_i = 500 + 280\,i$
and $i\in[0,1,\dots, 11]$.
This is $40\times12=480$ values of $\beta$ altogether. More than $4\times10^6$
energy levels were computed for each $\epsilon$. The size of the intervals
in $k$ was chosen to be maximal and such that the Brody distribution gives
a good fit to the level spacing distributions of the levels in the intervals,
meaning that $\bet$ is well defined.

For each $\beta(\epsilon_j,(k_i,k_{i+1}))$ an associated
localization measure $\mA$
was computed on a sample of 1000 consecutive levels around
$\bar{k}_i = (k_i + k_{i+1})/2$,
which is a mean value of $k$ on the interval $(k_i, k_{i+1})$.
Moreover, the obtained distribution functions $P(A)$ were calculated
for 22 values of $\ep$ and 12 values of $\bar{k}$, and some selection
of them is presented and discussed in the previous section \ref{sec4}.

\begin{figure}[H]
  \centering
  \includegraphics[width=9cm]{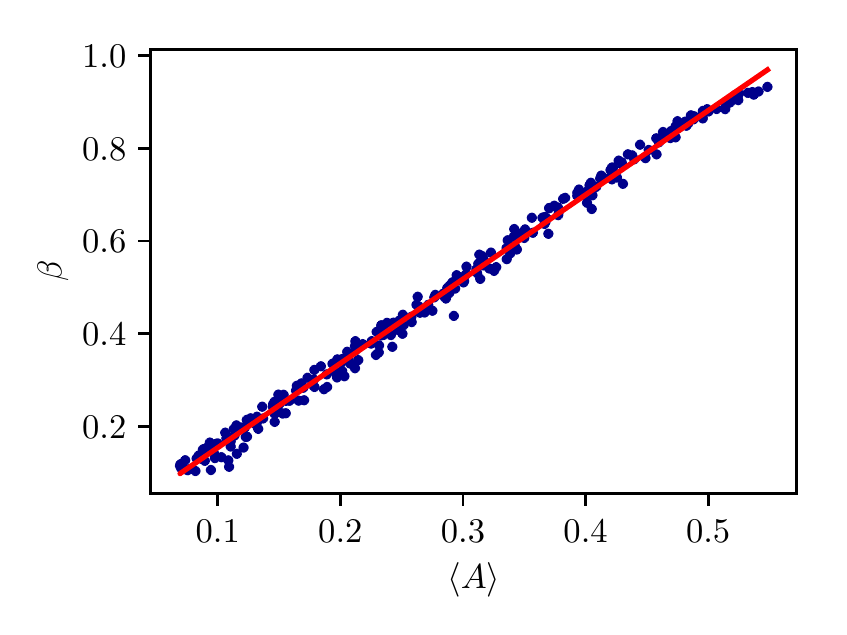}
  \caption{The level repulsion exponent $\beta$ as a function of the
    entropy localization measure $\mA$ for variety of stadia of
    different shapes $\ep$ and energies $E=k^2$, as defined in the text.}
  \label{betaVsA}
\end{figure}

For completeness, the almost linear dependence of $\beta$ on $\mA$,
obtained in \cite{BLR2018}, is shown  in Fig. \ref{betaVsA}.

This relation $\bet(\mA)$ is very similar to the
case of the quantum kicked rotator
\cite{Izr1990,ManRob2013,BatManRob2013}. In both cases the scattering
of points around the mean linear behaviour is significant, and
it is related to the fact that the localization measure $A$ 
of eigenstates has some distribution $P(A)$, as observed and discussed in
Ref. \cite{ManRob2015} for the quantum kicked rotator, and discussed
for the stadium billiard in the previous section \ref{sec4}.

There is still a great lack in theoretical understanding of the 
physical origin of this phenomenon,
even in the case of (the long standing research on) 
the quantum kicked rotator, 
except for the intuitive idea, that energy spectral properties should be 
only a function of the degree of localization, because the localization
gradually decouples the energy eigenstates and levels, switching the linear
level repulsion $\beta=1$ (extendedness) to a power law
level repulsion with  exponent $\beta < 1$ (localization). 
The full physical explanation is open for the future.

\begin{figure}[H]
  \centering
  \includegraphics[width=9cm]{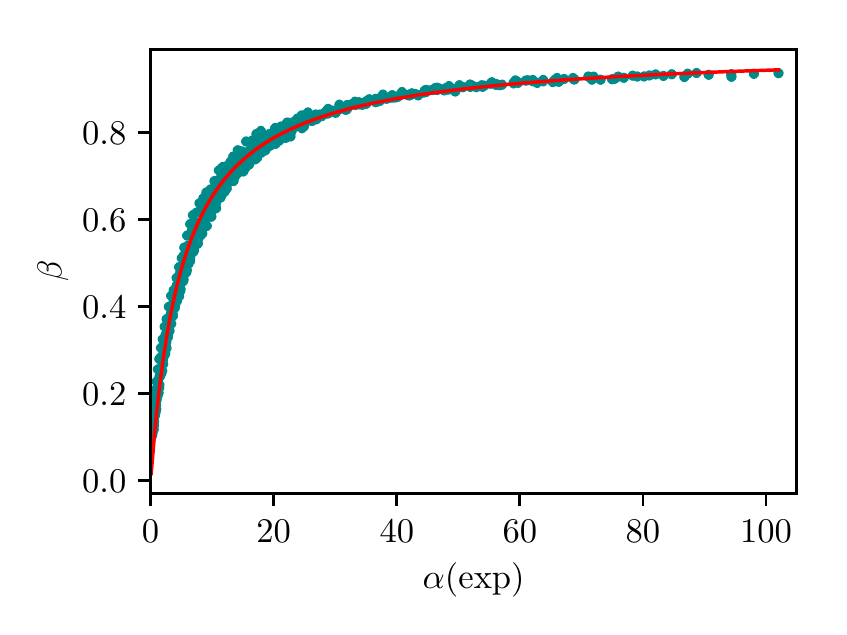}
  \caption{The level repulsion exponent $\beta$ as a function of $\al$ fitted
    by the function (\ref{BvsAExp}), based on $N_T$ from the exponential
    diffusion law. $\beta_{\infty}=0.98$ and $s=0.20$.}  
  \label{betavsalphaExp}
\end{figure}

\begin{figure}[H]
  \centering
  \includegraphics[width=9cm]{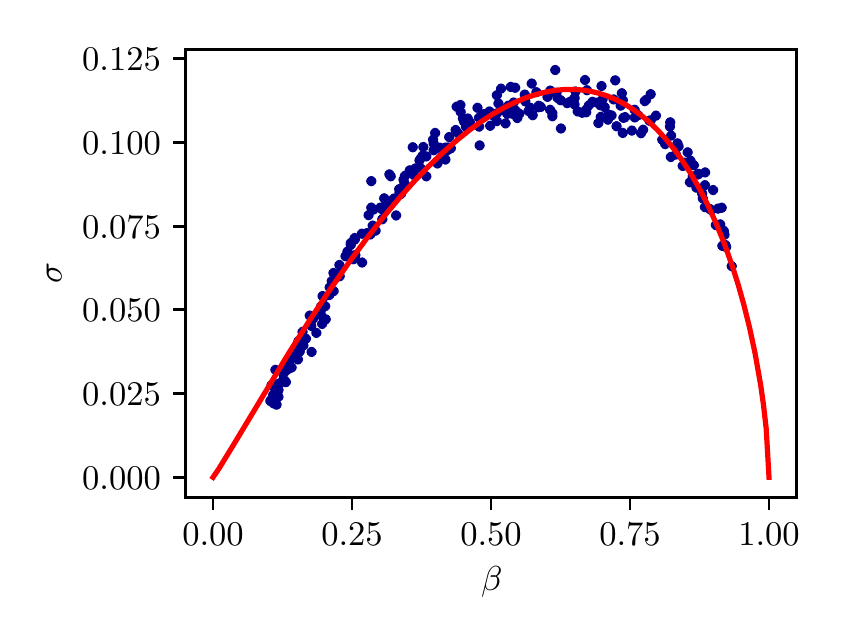}
  \caption{The standard deviation $\sigma$ as a function
    of the level repulsion exponent $\beta$ for variety of stadia of
    different shapes $\ep$ and energies $E=k^2$, as defined in the text.
    The best fitting curve for this {\em empirical function}
    is from Eq. (\ref{betadistr}), with the parameter values
    $C=0.345$, $a=1.07$ and $b=0.60$. $C$ is determined by fitting, not by
  normalization, as $\sigma(\bet)$ is a function, not a distribution.}
  \label{SigmaofBeta}
\end{figure}

As shown in \cite{BLR2018} and in Fig. \ref{betavsalphaExp}
the functional dependence of $\bet (\al)$ is always the rational function

\be  \label{BvsAExp}
\beta = \beta_{\infty} \frac{s\al}{1 +s \al}.
\ee
only the coefficient $s$ depends on the definition of $N_T$ and $\al$.
For the exponential law we found  $\beta_{\infty}=0.98$ and $s=0.20$.
Finally, we look at the dependence of $\sigma$ on $\bet$, shown in
Fig. \ref{SigmaofBeta}. It should be noted that dependence of $\sigma$ on
$\mA$ is nearly the same, because $\bet$ and $\mA$ are linearly related, as
demonstrated in Fig. \ref{betaVsA}.
It is surprising that also here the best fit of the function is
found by choosing the beta distribution (\ref{betadistr})
with the parameters $a=1.07$, $b=0.60$, $C=0.08$ and
$\bet$ is on the interval $[0,1]$.

\section{Conclusions  and discussion}
\label{sec6}

We have shown that in the stadium billiard of Bunimovich \cite{Bun1979}
the localization measure $A$, based on the
information entropy, of eigenstates as described by the
Poincar\'e - Husimi functions, has a distribution $P(A)$ very well described by
the beta distribution. We have also looked at the mean $\mA$ and the standard deviation
$\sigma$ as functions of the major control parameter $\al$, the ratio of the
Heisenberg time and the classical transport (diffusion) time.
We have also shown that the normalized inverse
participation ratio is equivalent to $A$.

The spectral level repulsion exponent $\beta$ of the localized
eigenstates is functionally related to $\mA$ \cite{BLR2018}.
Moreover, the dependence is linear, as
in the quantum kicked rotator, but somewhat different from the case of a mixed type billiard
studied recently by Batisti\'c and Robnik \cite{BatRob2013A,BatRob2013B}, where the
high-lying localized chaotic eigenstates have been analyzed after the separation of
regular and chaotic eigenstates.

$\beta$ is empirically a rational function of the
major control parameter $\al$. The definition of the classical transport
time is to some extent arbitrary, but we have shown in \cite{BLR2018}
that the various definitions do not change the shape
of the dependence on $\ep$, but instead affect
only the prefactor. As a consequence of that the dependence is
empirically (by best fit) always
a rational function. The transition from complete localization $\beta=0$
to the complete extendedness (delocalization) $\beta\approx 1$
takes place very smoothly, over about almost two decades of the parameter $\al$.

Thus we have again demonstrated by numerical
calculation that the fractional power law level repulsion
with the exponent $\beta \in [0,1]$ is manifested in localized chaotic 
eigenstates. The dependence $\bet (\mA)$ has some scatter due to the fact that
$A$ has above mentioned distribution $P(A)$ with nonzero $\sigma$.

Our empirical findings call for theoretical explanation, which is
a long standing open problem even for the main paradigm of quantum chaos,
the quantum kicked rotator studied extensively over the decades
\cite{Izr1990}.

Further theoretical work is in progress. Beyond the billiard systems,
there are many important applications in various physical systems,
like e.g. in hydrogen atom in strong magnetic
field \cite{Rob1981,Rob1982,HRW1989,WF1989,RWHG1994},
which is a paradigm of stationary quantum chaos,
or e.g. in microwave resonators, the experiments introduced by St\"ockmann
around 1990 and intensely further developed since then \cite{Stoe}.

\section{Acknowledgement}

This work was supported by the Slovenian Research Agency (ARRS) under
the grant J1-9112.

\bibliography{qChaos.bib}

\end{document}